\documentclass[aps, pra, a4paper, twocolumn, 10pt, showpacs]{revtex4-1}
\usepackage{amsmath, amssymb, amsthm, bm, textcomp}
\usepackage[T1]{fontenc}
\usepackage[latin9]{inputenc}
\usepackage{graphicx,graphics}
\usepackage{geometry}
\usepackage{color}
\newcommand{\be}{\begin{equation}}
\newcommand{\ee}{\end{equation}}
\newcommand{\eea}{\end{eqnarray}}
\newcommand{\bea}{\begin{eqnarray}}

\begin{document}

\title{Measurement-based Quantum Communication}

\author{M. Zwerger$^1$, H. J. Briegel$^{1}$ and W.\ D\"ur$^1$}

\affiliation{$^1$ Institut f\"ur Theoretische Physik, Universit\"at Innsbruck, Technikerstr. 21a, A-6020 Innsbruck,  Austria.}
\date{\today}

\begin{abstract}
We review and discuss the potential of using measurement-based elements in quantum communication schemes, where certain tasks are realized with the help of entangled resource states that are processed by measurements. We consider long-range quantum communication based on the transmission of encoded quantum states, where encoding, decoding and syndrome read-out are implemented using small-scale resource states. We also discuss entanglement-based schemes and consider measurement-based quantum repeaters. An important element in these schemes is entanglement purification, which can also be implemented in a measurement-based way. We analyze the influence of noise and imperfections in these schemes, and show that measurement-based implementation allows for very large error thresholds of the order of 10\% noise per qubit and more. We show how to obtain optimal resource states for different tasks, and discuss first experimental realizations of measurement-based quantum error correction using trapped ions and photons.
\end{abstract}
%

\maketitle


\section{Introduction}
Quantum communication is one of the most promising applications of quantum science, which is already at the edge of being used in practice and commercially. Quantum cryptography \cite{Sca09} is arguably the key application in this context, where the transmission of quantum states allows one to establish a secret cryptographic key between two communication partners, enabling them to communicate securely. Security is guaranteed by the laws of quantum mechanics, rather than by (unproven) assumptions on the computational complexity of certain tasks, as in classical encryption schemes. Quantum communication is however also important in the context of distributed quantum computation, where one may envision either a central quantum computation station that accepts quantum inputs from external nodes, or quantum computation that is performed by a network of small-scale processors that are connected via quantum channels. Of particular interest in this context is the method of blind quantum computation which can realized in a measurement-based setup \cite{Bro09,Bar12}. Other applications of distributed entangled states, such as secret sharing or secret voting, have been discussed. For all these applications of a quantum communication network, the term quantum internet has been suggested \cite{Kimble08}, in analogy with the classical internet.

The key task is to enable two communication partners, say Alice and Bob, to exchange quantum states and thus to transmit quantum information. The quantum channel connecting Alice and Bob will in general be noisy, and one needs to establish schemes to deal with this noise. This is particularly challenging for long-range communication, where novel strategies extending the classical approaches are required. Quantum information cannot be cloned or amplified, which implies that classical repeater schemes where signals are amplified and refreshed at regularly spaced intervals are not applicable. Different schemes have been invented to overcome this problem, and in principle allow one to establish quantum communication over arbitrary distances with only polynomial overhead. One is based on the direct transmission of encoded quantum information, where quantum error correction is repeatedly applied \cite{KL96}. A second approach has become known as the (entanglement-based) quantum repeater, where long-distance entangled states are generated by a combination of entanglement swapping and entanglement purification \cite{Br98}. These pairs are then used to transmit arbitrary quantum information by means of teleportation \cite{Be93}. Many variants and combinations of these schemes have been put forward and analyzed \cite{Gi11}. Also satellite-based quantum communication is being actively pursued \cite{Je13}.

In the standard approach to long-range quantum communication, quantum error correction \cite{Sho95,Ste96,Go97}, entanglement swapping \cite{Be93,Zu93}, and entanglement purification \cite{Be96,De96,Du07} are realized by implementing gates and operations. Here we discuss the potential of realizing parts of these schemes using concepts and techniques from measurement-based quantum information processing. In this approach, certain entangled states serve as a resource that allow one to implement different (sub-)tasks solely by means of measurements. We thereby take the concept of a measurement-based quantum computer \cite{Ra01} as a prototype model, where e.g. in the one-way model a large entangled resource state, the 2D cluster state \cite{Ra01b}, serves as a resource to realize {\em any} quantum operation. We concentrate on the implementation of specific tasks using small-scale resources of minimal size -- only containing input and output particles--, and show how to identify such resources. We demonstrate that one can realize circuits that comprise a large number of gates in a compact way, where the reduced number of resources leads to a significant improvement with respect to noise tolerance of the schemes. We find, in fact, that for various tasks the error threshold can be 10\% per qubit or even higher, which makes such schemes very attractive from an experimental perspective.

The paper provides a review of recent results published in \cite{Zw12,Zw13,Zw14a,La13,Ba14,Zw14b} together with additional discussions and observations, and is organized as follows. In Sec. \ref{Sec_Background} we provide the necessary background on quantum communication schemes and measurement-based information processing. We also discuss the construction of optimal resource states of minimal size for different tasks using the Choi-Jamiolkowski \cite{Ja72,Ch75} isomorphism, and introduce error models to describe noise and imperfections. In Sec. \ref{Sec_EPP} we consider measurement-based entanglement purification, and show a universal error threshold of 24\% noise per particle. We also discuss measurement-based implementations of hashing protocols, which, remarkably and in stark contrast to a gate-based implementation, are practical and represent a viable option. We then consider measurement-based quantum error correction in Sec. \ref{Sec_ErrorCorrection}, and describe resource states for encoding, syndrome read-out and decoding for different error correction codes. We also derive error thresholds for the applicability of the schemes. In Sec. \ref{Sec_EncodedQC} we discuss direct communication of encoded quantum information using quantum error correction, while in Sec. \ref{Sec_Repeaters} we turn to entanglement-based long-range quantum communication using quantum repeaters. In Sec. \ref{Sec_Hybrid} we describe an extension of the approach to a hybrid quantum computation scheme, and discuss first experimental implementations using trapped ions and photons in Sec. \ref{Sec_Experiments}. In Sec. \ref{Sec_Other} we briefly comment on some alternative schemes that make use of measurement-based elements. We summarize and conclude in Sec. \ref{Sec_Summary}.

\section{Background}
\label{Sec_Background}
In this section we provide background information on schemes for long-range quantum communication, quantum error correction and measurement-based quantum information processing. We only give a brief overview, and refer the reader to review articles on the different subjects \cite{Go09,Ra12,Du07,Gi11,Br09} 

\subsection{Quantum error correction}
Quantum error correction \cite{Sho95,Ste96,Go97} is an extension of the redundant encoding of classical information. The basic idea is to use several physical qubits to encode one logical qubit, in such a way that errors can be detected and corrected. As quantum states can be in a superposition of basis states, it is not sufficient to correct errors only on basis states, but it is required that a whole subspace is protected. For a single logical qubit, this can be achieved as follows: (i) One specifies the kind of errors one wants the code to correct and identifies a basis for error operators. For example, the Pauli operators represent a basis for all possible error operators that can act on a single qubit. Consequently, for a code that should protect against arbitrary errors occurring on a single system, the relevant error operators are given by $S_i \in \{ \sigma_\alpha^{(k)} \}$. (ii) One chooses a two-dimensional subspace $P_0$ in such a way that each error operator maps the subspace to an orthogonal two-dimensional subspace, $P_i = S_i P_0 S_i^\dagger$ with $P_i P_j = \delta _{ij} P_i$.; (iii) One chooses a basis in the logical subspace $P_0$ and defines logical basis states $|0_L\rangle, |1_L\rangle$. (iv) Error correction is done by projecting the system onto the subspaces $\{P_j\}$, where for result $j$ the error operator $S_j^\dagger$ is applied to recover the quantum information. Notice that the measurement leads to a digitalization of the error. This implies that quantum error correction can deal also with combination of errors, and it is sufficient to consider only a basis of error operators.

Constructions for good and efficient error correction codes are known. The simplest code that allows one to protect against bit flip errors is a repetition code with $|0_L\rangle = |0\rangle^{\otimes m}, |1_L\rangle = |1\rangle^{\otimes m}$. For $m=2n+1$, such a code can correct for up to $n$ bit flip errors. Similarly, a code to protect against phase flip errors can be obtained by applying a Hadamard operation on each qubit. A code that can protect against an arbitrary error happening on a single qubit has a minimal size of 5, i.e. 5 qubits are required to to form a logical qubit. Codewords of error correcting codes corresponding to bit-flip errors and arbitrary errors are depicted in Fig. \ref{Fig_Codes}.

We use graph states \cite{He04,He06} to describe the codewords. For any graph $G=(V,E)$ with vertices $k \in V$ and edges $(k,l) \in E$, the corresponding graph state is given by
\be
|\psi_G\rangle = \prod_{(k,l) \in E} U_{\rm PG}^{(kl)} |+\rangle^{\otimes N},
\ee
where $U_{\rm PG}=diag(1,1,1,-1)$ and $|\pm\rangle =\tfrac{1}{\sqrt{2}}(|0\rangle \pm |1\rangle)$. We use the graph to describe the entanglement structure of the state, in many cases throughout the article there will be additional local unitary operations that are required in addition to obtain the actual state in question.

\subsection{Quantum communication based on quantum error correction}
\label{SecKL}
The first scheme for scalable quantum communication was put forward in 1996 and is based on the transmission of encoded quantum information. It works as follows \cite{KL96}. The quantum information to be transmitted is encoded at Alice's site using some quantum error correction code, where $N$ physical qubits are used to encode one logical qubit. The long channel is divided into small segments, and after each segment an error correction step (i.e. syndrome read out and if required correction operation) are performed. At the final station, quantum information is decoded.

A polynomial scaling of the resources with the distance was derived analytically \cite{KL96}. Any stabilizer quantum error correction code is suitable for such a procedure. The error threshold was derived for a gate-based implementation, and was found to be of the order of $10^{-5}$ for the involved gates - basically the same as for universal fault-tolerant quantum computation. Notice that these initial results have been significantly improved  meanwhile, as advances in fault-tolerant quantum computation are directly applicable also in such a communication scenario. The scaling for required resources was found to be poly-logarithmic \cite{NiCh98}, and the threshold value for fault-tolerant quantum computation -and hence encoded quantum communication- is about $10^{-2}$ \cite{Kn05,Rau07}.

\subsection{Entanglement purification}
\label{SecEPP}
Entanglement purification is an important primitive in quantum communication \cite{Be96,De96,Du07}. Consider two parties A and B that are connected by a noisy quantum channel and whose aim is to establish a maximally entangled state $|\phi^+\rangle = \tfrac{1}{\sqrt{2}}(|00\rangle + |11\rangle)$. A straightforward approach would be to produce such a state locally, i.e., at the location of one party, and send one qubit of the state through the noisy channel. Due to noise and imperfections, the parties will end up with a noisy entangled state instead. They can, however, repeat the procedure and produce in this way many copies of such noisy entangled states. Entanglement purification then allows the parties to create out of many such noisy pairs solely by performing local operations on their system, together with classical communication, fewer pairs with an increased fidelity, that is, with a smaller amount of noise. These pairs can then be used for teleportation, and hence to reliably transmit arbitrary quantum information.

One important entanglement purification protocol is the recurrence protocol of \cite{De96}, which operates always on two copies and is applied in an iterative way. At each step, two identical copies resulting from the previous purification step are processed, and the fidelity is iteratively increased. The basic idea is to concentrate the entanglement of two pairs into a single one, or alternatively one can say that one attempts to learn (non-local) information about the first pair with help of the second pair. This increase in information on the first pair is equivalent to a larger fidelity.
Each purification step consists of the application of local, joint operations on the two pairs, followed by a measurement of the second pair. Depending on the measurement outcome, the remaining pair is kept or discarded. Hence purification only works probabilistically, however the procedure quickly converges to unit fidelity in the noiseless case, provided the initial fidelity of the pairs is sufficiently large, $F \geq 1/2$. Taking noise and imperfections in local operations and measurements into account, the purification regime gets smaller, i.e. only some maximal fidelity smaller than unity can be reached, and a larger initial fidelity is required. The threshold for local noise is determined by the existence of a non-vanishing purification regime, i.e. a range of possible initial values of $F$ for which the fidelity can be increased. Depending on the error model used, the acceptable noise per operation is given by a few percent \cite{Du07}. Alternative entanglement purification protocols that operate simultaneously on a larger number of input pairs also exist \cite{Be96a,As04}. One example of such a protocol is the hashing protocol \cite{Be96a}, which operates on a very large number of copies $N$, and produces $M < N$ almost perfect pairs. Again, the main idea of the protocol is to measure $N-M$ pairs to obtain information about the remaining ensemble. In fact, measuring $N S(\rho_A)$ pairs is sufficient to be left with a pure-state ensemble of maximally entangled pairs, where $S(\rho_A)$ is the Von-Neumann entropy of the reduced state of a pair. While hashing has --in contrast to the recurrence protocol-- a non-zero yield, it is not applicable under non-idealized conditions \cite{Du07}. Since the operations that need to be applied repeatedly at each of the pairs are noisy, that noise accumulates, thereby jeopardizing the entanglement purification effect \cite{Du07}.

\subsection{Quantum repeaters}
\label{Sec_Repeater}
For long-distance quantum communication, entanglement purification alone is not sufficient. Clearly, the initial state still needs to be entangled, i.e. a large enough initial fidelity, in the case of perfect local operations larger than $1/2$, is required. However, losses and noise increase exponentially with the distance, thereby limiting the maximal distance to a few hundred kilometers when using photons transmitted through optical fibers. For larger distances, other methods are required. The quantum repeater \cite{Br98} is a scheme that allows for efficient long-distance quantum communication. It utilizes a nested combination of entanglement purification and entanglement swapping, thereby generating high fidelity entangled pairs that can then be used to teleport arbitrary quantum information. The basic idea is to split the long channel into smaller segments, and generate short-distance entangled pairs that are purified to some working fidelity $F_0$. These short-distance pairs are then connected via Bell measurements, a process that has been termed entanglement swapping, to form an entangled pair of longer distance. For non-ideal pairs or noisy operations, the fidelity of the resulting pairs are reduced, and only a few can be connected in this way. One then uses entanglement purification to re-purify these mid-distance pairs to the working fidelity $F_0$. When applied to all segments simultaneously, this leaves us with the same situation as initially, except that the distance of the pairs is enlarged. This procedure can than be applied in a nested way, thereby always at least doubling the distance. The required resources, i.e. the number of elementary pairs, can be shown to increase only polynomial with the distance \cite{Br98}.

Many variants of this scheme that use different methods to purify the pairs have been put forward, for a recent review see e.g. \cite{Gi11}. They allow one to reduce the number of parallel channels significantly, and also to deal, up to certain distances, with memory errors \cite{Ha07}. Similar as in entanglement purification, noise in local operation limits the applicability of the scheme, and one finds a noise threshold that is a bit lower than the one for entanglement purification alone, but still of the order of a few percent \cite{Br98}. This is several orders of magnitude larger than for the original scheme based on quantum error correction \cite{KL96} (see also Sec. \ref{SecKL}) and consequently most of the research activities have focused on the (entanglement-based) quantum repeater.

\subsection{Measurement-based quantum information processing}
Here we describe how gates and operations can be realized in a measurement-based way using certain highly entangled resource states. This approach, initially put forward in the context of quantum computation, differs significantly from the standard gate-based approach. One does not need to apply coherent operations on many systems, but rather prepare certain entangled resource states by some means. This preparation procedure can even be probabilistic, and may in many systems be easier than performing coherent gates \cite{Nie04,ProbGates}. For example, achieving interactions between the polarization degree of freedom of two photons in a deterministic way is a very challenging task, while the probabilistic preparation of entangled photon states, e.g., by means of parametric down conversion, is routinely done in many laboratories. In this sense, measurement-based quantum information processing offers a potential experimental advantage. We will strengthen this claim by deriving error thresholds for different tasks, where we find that they can be significantly increased by making the resource states smaller. A direct comparison with a gate-based implementation is however difficult, as different error models are used.

A well-studied measurement-based model is the one-way quantum computer \cite{Ra01}, where a so-called 2D cluster state \cite{Ra01b} serves as a universal resource for quantum computation. Solely by means of single-qubit measurements, quantum information is processed, where the choice of the measurement bases determines the gates or the algorithm that is implemented. In general, these measurements need to be done sequentially in an adaptive way. An important exception are gates and circuits from the Clifford group, where all measurements can be done in a parallel fashion \cite{Ra03}. The reason for this is that Pauli byproduct operators due to random measurement outcomes can be commuted through all gates of this kind, and be executed at the end of a given quantum circuit simulated on the 2D cluster. Equivalently, one can also use more compact resource states of smaller size to implement a specific circuit. In the case of Clifford circuits, these resource states consist only of input and output qubits. In the following we will discuss how to construct such resource states of minimal size.

\subsection{Construction of resource states}
\label{Sec_Resources}
Starting from a 2D cluster states as universal resource, each gate of a circuit ${\cal C}$ is associated with a specific measurement pattern \cite{Ra01}. For a Clifford Circuit, all measurements, except the ones on input particles, can be performed beforehand. In this case, Pauli measurements suffice, and the resulting special purpose resource state is --up to local correction operations-- a graph state can be determined by update rules \cite{He04}.

An alternative way of obtaining the resource states is given by the Choi-Jamiolkowski isomorphism \cite{Ja72,Ch75}. For a $N$ qubit unitary operation $U$, the corresponding resource state $|\psi_U\rangle$ is given by
\be
|\psi_U\rangle = (I \otimes U)  \left( \tfrac{1}{2^{N/2}}\sum_{\bm k= \bm 0}^{\bm 2^{N}-1} |{\bm k}\rangle_A \otimes |{\bm k}\rangle_B \right),
\ee
where we use binary notation to denote the $N$-qubit state $|{\bm k}\rangle$. That is, $U$ acts on part of a maximally entangled state of $2N$ qubits. In turn, $|\psi_U\rangle$ allows one to implement $U$ on an arbitrary $N$-qubit input state $|\varphi\rangle$ probabilistically. Consider the Bell basis $\{|\phi_i\rangle=I \otimes \sigma_i^{*}|\phi_0\rangle\}$ with $|\phi_0\rangle = |\phi^+\rangle = \tfrac{1}{\sqrt{2}}(|00\rangle +  |11\rangle)$. On qubit $k$ of the input state and part $A$ of the state $|\psi_U\rangle$, a Bell measurement is performed and outcome $i_k$ is obtained. It is easy to check that the resulting state at site $B$ is then given by
\be
U (\sigma_{i_1}\otimes \sigma_{i_2} \ldots \otimes \sigma_{i_N})|\varphi\rangle.
\ee
If all measurement outcomes were $0$, then the desired operation has been performed. Notice that in general it is not possible to commute Pauli operations through the unitary without altering $U$. This would be necessary to later apply a correction operation, and thus for general $U$ this procedure is necessarily probabilistic. For a deterministic implementation, a larger resource state with additional auxiliary particles that are measured sequentially and in an adaptive way, would be required. However, if $U=U_C$ is a Clifford gate or a Clifford circuit, i.e. an (arbitrary long) quantum circuit that consists solely of Clifford gates, then by definition $U_C$ transforms tensor products of Pauli operators into (possibly different) tensor products of Pauli operators. One thus obtains
\be
(\sigma_{j_1}\otimes \sigma_{j_2} \ldots \otimes \sigma_{j_N})U_C|\varphi\rangle,
\ee
where ${\bm j}=f({\bm i})$ is some (known) function of the initial measurement outcomes. It follows that the knowledge of the measurement outcomes allows one to correct at site $B$ the byproduct operations by applying $(\sigma_{j_1}\otimes \sigma_{j_2} \ldots \otimes \sigma_{j_N})$, and hence to realize $U_C$ deterministically to an arbitrary input state.

In some cases, circuits may also include measurements on output particles. If these measurements are of Pauli type, then one can actually perform them beforehand. That is, one takes the resource state, applies the appropriate projection, where one assumes that the measurement outcome is $+1$, and obtains in this way a reduced state where the to-be-measured particle is no longer included. The actual measurement outcome can be determined from the results of the in-coupling Bell measurements. This is possible because all measurements, including the teleportation-like process to implement the gate or circuit, take place on different particles and hence commute. In addition, the correction operations on output particles, also the ones to be measured at a later stage, are only Pauli operations. This simply leads to a re-interpretation of measurement outcomes. This is explained in detail for circuits corresponding to entanglement purification in \cite{Zw12}, and for error correction circuits in \cite{Zw14a} (see also \cite{He06}).
The resource states are in fact all stabilizer states, local unitary equivalent to graph states. This follows from the alternative construction using the 2D cluster state, but can also be checked directly by using a stabilizer description of the resource states.

All circuits we will consider in this work are of Clifford type, so the above construction provides us with resource states of minimal size containing only input and output particles. This is true for encoding and decoding circuits, error syndrome readout, entanglement swapping as well as entanglement purification. All auxiliary particles that are eventually measured in a circuit-based implementation are not required in this measurement-based approach. In addition, the size of the resource state does not depend on the complexity of the Clifford circuit, i.e. even a complex circuit that contains many gates does only require input and output particles, only the structure of the entangled state changes. Notice that this reduction of size of resource states is {\em the} crucial feature that makes measurement-based implementation of such circuits very attractive, and leads to high error thresholds as shown below.

\subsection{Combination of resource states}
\label{Sec_Combination}
We remark that the construction of resource states for a combination of tasks is straightforward, and can be done by combining the corresponding states in a way to be described. Consider for instance the implementation of an encoded Clifford gate, e.g. a Hadamard gate, with built-in error correction. If the resource states for encoding, syndrome-readout, decoding and for the non-encoded single qubit gate are known, one can simply connect the corresponding input and output particles by Bell measurements. These measurements are also of Clifford type, and can hence be done beforehand. That is, one uses resource states for (i) decoding; (ii) unencoded gate; (iii) encoding; (iv) syndrome read-out and combines them to form a $2N$ qubit resource state for the combined task.

Similarly, one obtains a combined state of $N+2$ qubits that allows for encoding, syndrome read out and decoding, as is illustrated in Fig. \ref{Fig_Codes}.

\subsection{Error model}
In any realistic situation, noise and imperfection will occur and limit the performance of any quantum information processing protocol. In a measurement-based implementation, the only sources of noise are (i) imperfect resource states and (ii) noisy measurements. We will use a simple, heuristic model to describe both kinds of errors. We will discuss later in which sense this model is justified, and that it can in fact be actively ``enforced''.

We consider a completely positive map that describes depolarization noise, sometimes also called white noise, acting on particle $a$ by
\be
\label{mapnoise}
{\cal E}_a(p)\rho= p \rho + (1-p) \tfrac{1}{4} \sum_{j=0}^3 \sigma_j^{(a)} \rho \sigma_j^{(a)},
\ee
where $p$ is the error parameter. That is, with probability $p$ the particle remains unaltered, while it is completely depolarized with probability $(1-p)$.
Noisy measurements can be described by a two-step process, where in the first step noise acts on all particles that are subjected to the measurement, and in a second step a perfect measurement is applied. In case of a Bell measurement ${\cal P}$ we have
\be
\label{noisyBell}
{\cal P}{\cal E}_a(q){\cal E}_b(q)
\ee
Noisy resource states are described by depolarizing noise acting on each of the particles of the resource state, i.e.
\be
\label{rho_U}
\rho_U(p) = \prod_{a=1}^N {\cal E}_a(p) |\psi_U\rangle\langle \psi_U|.
\ee
Notice that this error model takes into account that larger resource states are stronger affected by noise, as the fidelity drops (in general exponentially) with the size of the resource state $N$. It is heuristic in the sense that only uncorrelated noise is considered, and all resource states are affected in the same way, independent of the complexity of their creation. However, this is in fact a desired feature rather than a drawback. Since we are dealing with a measurement-based implementation, we do not wish to specify or restrict the way resource states are generated. It should be emphasized that all thresholds we provide in the remaining article are with respect to this error model, and a direct comparison with other, e.g. gate-based error models, is not possible.

In fact, many possibilities to generate highly entangled states exist. The standard approach consists of applying sequences of gates to some initial state, usually a product state. However, the set of gates that is available, as well as their quality, strongly depends on the the particular set-up. This makes the judgement of how complicate the preparation of a given state is ambiguous. For instance, usually the preparation of a so-called GHZ state $\tfrac{1}{\sqrt 2}(|0\rangle^{\otimes N} + |1\rangle^{\otimes N})$ requires $N-1$ two-qubit CNOT gates, while in present ion-trap set-ups a single interaction, corresponding to a joint $N$ qubit gate, suffices \cite{MS99}. Other ways of state preparation are conceivable, e.g. by cooling of a strongly interacting system to its ground state, or by reservoir engineering \cite{Di08,Ve09}.

In addition, probabilistic methods for resource state preparation are available, as resource states only enter in the quantum information processing process once they have been prepared \cite{Nie04,ProbGates}. In contrast, a gate-based implementation requires deterministic gates to not jeopardize quantum information processing. This also offers advantages and new possibilities in certain set-ups. Consider e.g. a photonic implementation, where quantum information is stored in the polarization degree of freedom of different photons. Obtaining a deterministic, controlled interaction is very hard and demanding, and the used non-linearities are orders of magnitudes too weak to obtain a direct two-qubit gate. The probabilistic preparation of certain entangled resource states is however possible, and is in fact a standard procedure in many labs. For instance, parametric down conversion allows one to generate entangled states of moderate size. The possibility of generating resource states probabilistically also opens the way to use tools such as entanglement purification to generate resource state with high fidelity. That is, whenever resource states can be generated by some means, even with low fidelity, one can use entanglement purification to increase the fidelity and hence reduce errors. As all resource states we consider throughout this article are local unitary equivalent to graph states, multipartite entanglement purification protocols exist \cite{Du03,Gl06,Kr06}.

An additional feature of entanglement purification is that the shape of noise will be modified. In fact, first results indicate that entanglement purification will bring the state closer to the form Eq. \ref{rho_U}, i.e. to local noise. The reason is that noise due to imperfect entanglement purification enters locally, while the process as such reduces all kinds of noise. This is currently under investigation and will be reported elsewhere \cite{Wal15}.

We also would like to remark that when performing a Bell measurement on a qubit that is affected by noise described by Eq. \ref{mapnoise} (or more generally by any Pauli-diagonal noise process), one can move the action of noise onto the other qubit. That is ${\cal P}_{ab} {\cal E}_a(p) \rho = {\cal P}_{ab} {\cal E}_b(p) \rho$, where ${\cal P}$ denotes the Bell measurement. This can be shown by considering the corresponding Choi-Jamiolkowski states of the two processes \cite{Zw13}. This will prove to be a very powerful tool in the analysis of error thresholds. Notice also that ${\cal E}(p_1){\cal E}(p_2) \rho = {\cal E}(p_1p_2) \rho$, which can again be verified by direct computation. When considering noisy Bell measurements, described by Eq. \ref{noisyBell}, and noisy resource states described by Eq. \ref{rho_U}, one can summarize the effect of all imperfections by a single noise channel ${\cal E}(q^2p)$ acting on one of the particles where the Bell measurement is performed. We will hence in the following set $q=1$, i.e. consider perfect Bell measurement, as the effect of a noisy Bell measurement is the same as for noisy resource states with a modified error parameter $p$. This error parameter should be understood as representing all kinds of imperfections in preparation and noisy Bell measurements.

\section{Measurement-based entanglement purification}
\label{Sec_EPP}
We now turn to measurement-based entanglement purification, where we analyze different purification protocols and their performance under noise.

\subsection{Recurrence protocol}
\begin{figure}[ht]
\begin{picture}(210,140)
\put(-20,-80){\includegraphics[keepaspectratio, width=10cm]{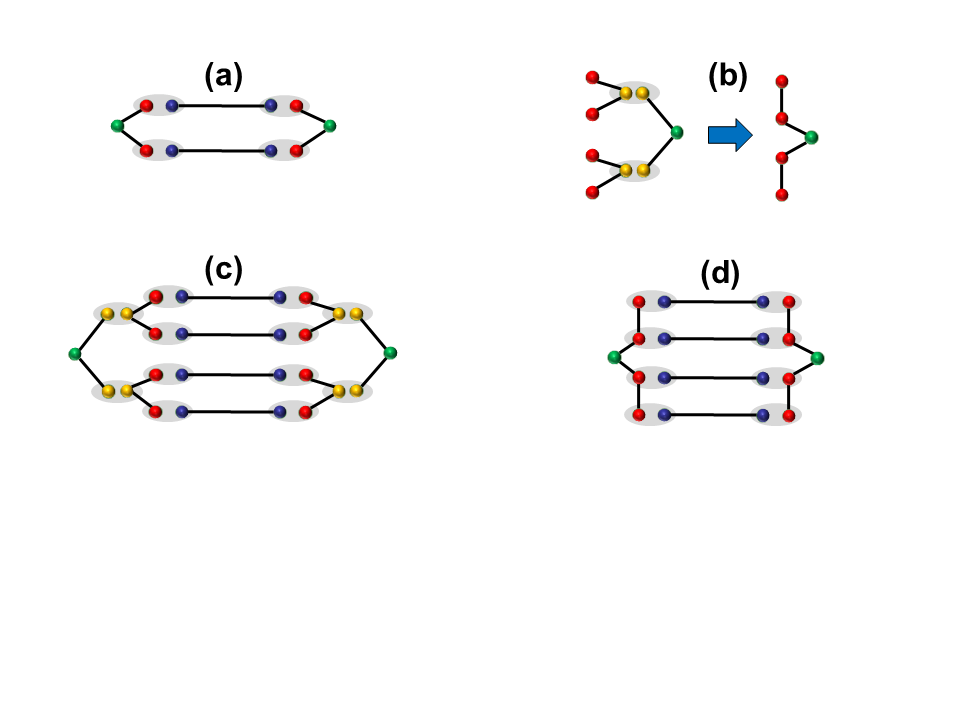}}
\end{picture}
\caption{Illustration of measurement-based entanglement purification. Blue particles correspond to entangled pairs shared between Alice and Bob, the remaining particles belong to resource states where red particles are input states, blue particles output states and particles that are virtual or can be removed are depicted in yellow. Bell measurements are symbolized by shaded ellipses. (a) Single step of the $2 \to 1$ recurrence protocol. (b) Resource state at Bobs site for two steps of the recurrence protocol. Size of the resource state can be reduced to 5 particles when intermediate Bell-measurements are performed beforehand. (c) Two steps of the recurrence protocol, corresponding to a $4 \to 1$ protocol without and (d) with reduced size resource states.}
\label{Fig_EPP}
\end{figure}

We start by considering a single step of the $2 \to 1$ recurrence protocol described in Sec. \ref{Sec_EPP} (see also \cite{Zw12}) that probabilistically produces a single pair of higher fidelity from two initial pairs. This protocol can be implemented in a measurement-based way using a three-qubit state at each site, i.e. at Alice and Bob, that is LU equivalent to a GHZ state. Two of the qubits correspond to the input, while one corresponds to the output particle. This makes use of the fact that the measurement on the second output pair can be done beforehand, as described in Sec. \ref{Sec_Resources} \cite{Zw12}. Entanglement purification takes place by performing Bell measurements at Alice and Bob on the two pairs and the input particles of the resource states, see Fig. \ref{Fig_EPP}. Depending on the outcome of the Bell measurements, the resulting pair is discarded or kept (see \cite{Zw12} for conditions), and eventually a Pauli byproduct operation is applied.

The procedure can be applied in an iterative way, as shown in Fig \ref{Fig_EPP}, taking output pairs of the first step as input pairs for the second step. Notice that in this case both steps can be unified into a single step, thereby realizing a $4 \to 1$ protocol. Formally, this can be done by considering the two steps directly and constructing the resulting resource state, or by considering the resource states for the two-step procedure and performing the second Bell measurement that couples the resource states beforehand. This leads to a 5-qubit resource state (4 input and one output particle) at each site, as opposed to 9 particles when realizing the procedure sequentially. The resource state is LU equivalent to a linear cluster state, i.e. a graph state of a chain of 5 particles where the middle one corresponds to the output. However, the success probability as compared to a stepwise realization is reduced, as all three purification steps need to be successful simultaneously, while the step-wise procedure allows to combine only pairs from successful branches of the previous steps. In principle, one can iterate this procedure further an obtain resource states of size $2^m+1$ for $m$ purification rounds. Similarly, any $N \to M$ entanglement purification protocol \cite{Be96a,As04} simply corresponds to a $N+M$ qubit resource state.

\subsection{Universal error threshold}
We now analyze the performance of measurement-based entanglement purification in the presence of noise and imperfections. We consider noisy input states described by Eq. \ref{rho_U}. The error parameter $p$ describes the strength of noise per particle, and we refer to $1-p$ as noise level with e.g. $5\%$ noise corresponding to $p=0.95$.

It was observed in \cite{Zw13} that the error threshold for measurement-based entanglement purification is given by $3.5\%$, $7.1\%$ and $10.4\%$ for the $1 \to 2$, $1 \to 4$ and and $1 \to 8$ protocol respectively, indicating that the reduction of number of particles indeed gives an advantage. It is however possible to derive a universal and optimal error threshold for arbitrary $N \to 1$ protocols, as shown in \cite{Zw13}. To this aim we make use of the fact that for Bell measurements, noise can effectively be moved from one particle (input particle of resource state) to the other particle (particle of the pair to be purified). That is, the initial fidelity of the input pair is decreased.  Noise acting on the output particle of the resource state can be considered at the end of the process, at which stage it simply reduces the fidelity of the output pair. This leaves us with a {\em noiseless} resource state, corresponding to a perfect entanglement purification protocol that is capable of producing maximally entangled pairs with fidelity $F=1$. Independent of the entanglement protocol used, the minimal required fidelity of the initial pairs is give by $F > 1/2$.

The threshold can now be easily determined. The conditions are that (i) the initial fidelity is larger than $1/2$ (i.e. $p>1/3$), and (ii) the fidelity of the output pair, when taking noise of output particle of the resource state into account, should be larger than the one of the input pair. We describe the initial pair by ${\cal E}_a(q){\cal E}_b(q) |\phi^+\rangle\langle \phi^+| = {\cal E}_a(q^2)|\phi^+\rangle\langle \phi^+|$ with $|\phi^+\rangle=(|00\rangle + |11\rangle)/\sqrt{2}$. The two conditions now read
\bea
q^2p^2 > \tfrac{1}{3}&\hspace{0.5cm},\hspace{0.5cm}&  p^2 \geq q^2
\eea
which yields an error threshold $p_{\rm min}=1/\sqrt[4]{3} \approx 1-0.24$. That is, up to $24\%$ of noise per particle are acceptable. Notice that we have made no assumption on the entanglement purification protocol, except that the initial fidelity of the pairs needs to be larger than $1/2$. For states diagonal in the Bell basis, pairs with fidelity $F \leq 1/2$ are separable and hence not entangled. In this sense, the threshold is optimal and universal.

\subsection{Hashing protocol}
\label{Sec_Hashing}
We now consider $N \to M$ protocols, more specifically so-called hashing protocol. The hashing protocols uses bilateral CNOT operations, applied on a subset of pairs as source, and a specific pair as target, to reveal information about the source pairs by measuring the target pair \cite{Be96a}. All involved operations are of Clifford type, so one can construct a resource state of size $N+M$ that allows one to implement the protocol in a measurement-based way. In contrast to the recurrence protocol discussed before, hashing is a deterministic protocol. It is based solely on information gain about the remaining ensemble, where all remaining pairs except the pairs that are measured can be finally used. Consequently, hashing has a non-zero yield \cite{Be96a}.

We are interested in a measurement-based implementation, and in the effect of noise and imperfections. As demonstrated in \cite{Zw14b}, a measurement-based implementation is practical in the sense that about 7\% error per particle is acceptable in the resource states. The proof is very similar to the one presented for the universal error threshold, with the only difference that the required initial fidelity of the pairs for hashing is given by $F_{\rm min} \approx 0.8107$, yielding the slightly smaller error threshold.

It should be emphasized that {\em only} a measurement-based implementation makes hashing practical. In a gate-based approach, any non-zero gate error accumulates and renders hashing impossible \cite{Zw14b}. The key feature that makes the measurement-based implementation practical is the reduced size of the resource state. We remark that {\em any} $N \to M$ entanglement purification protocol \cite{Be96a,As04} that is based on Clifford operations (which is the case for all known protocols so far) can be implemented in a measurement-based way, with resource states of minimal size. In all cases, the error threshold is simply determined by the purification range of the noiseless protocol.

\section{Measurement-based quantum error correction}
\label{Sec_ErrorCorrection}
We now turn to quantum error correction, and analyze encoding, decoding and syndrome read-out. We consider the large class of stabilizer quantum error correction codes, which includes the Calderbank-Shor-Steane (CSS) error correction codes. For all these codes, codewords are stabilizer states, and the corresponding circuits for encoding, decoding and syndrome read-out are Clifford circuits. It follows that for a measurement-based implementation, resource states of minimal size involving only input and output particles, i.e. of size $N+1$ or $2N$, can be found \cite{Zw14a}. The error syndrome read-out and correction also requires auxiliary particles that are measured to reveal the error syndrome in a gate-based approach \cite{Sho95,Ste96,Go97}. As described above, in a measurement-based implementation one can measure these qubits beforehand and work with a resource state of size $2N$. The error syndrome and hence the required correction operations can be determined from the results of the in-coupling Bell measurements. This means that the advantage of a measurement-based implementation is twofold: on the one hand, one can replace a whole quantum circuit that involves many gates by a single resource state, independent of the size of the (Clifford) circuit. On the other hand, no auxiliary particles are required.

\subsection{Repetition code and five-qubit graph code}
\begin{figure}[ht]
\begin{picture}(210,225)
\put(-20,-140){\includegraphics[keepaspectratio, width=17cm]{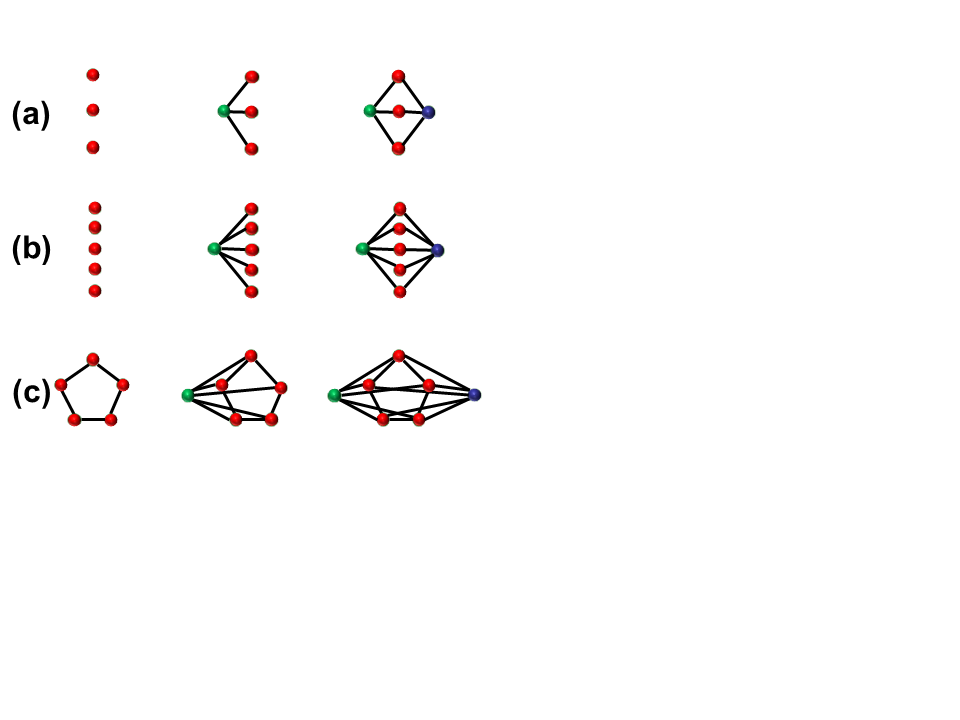}}
\end{picture}
\caption{Illustration of resource states for codewords (left column); encoding (middle column); and encoding, syndrome-read out, and decoding (right column) for (a) three-qubit repetition code, (b) five-qubit repetition code and (c) five-qubit ring code. All states are graph states up to local unitary operations.}
\label{Fig_Codes}
\end{figure}

We start by considering a simple $m=2n+1$ qubit repetition code that can correct for bit-flip errors occurring on up to $n$ particles. The codewords are given by $|0_L\rangle=|0\rangle^{\otimes m}, |1_L\rangle=|1\rangle^{\otimes m}$, and the resource states for encoding and decoding (with integrated error syndrome read-out) are given by GHZ-states $\tfrac{1}{\sqrt{2}}(|0\rangle_A|0_L\rangle_B + |1\rangle_A|1_L\rangle_B)$ (see Fig. \ref{Fig_Codes}). A combination of both procedures can be done with a $m+2$ qubit state \cite{La13}. For encoding, one starts with a qubit in some (possibly unknown) state $|\varphi\rangle = \alpha|0\rangle + \beta |1\rangle$, and performs a Bell measurement on this qubit and qubit $A$ of the resource state. This leaves us --up to a (logical) Pauli correction-- with an encoded state $|\varphi_L\rangle = \alpha |0_L\rangle + |1_L\rangle$ of system $B$. For decoding and syndrome read out, Bell measurements on all qubits of the encoded system and system $B$ of the resource state are performed. The decoded quantum information is stored in $A$, where the required correction operation can be determined from the results of the Bell measurements as described in \cite{Zw14a}.

In a similar way, one can consider the 5-qubit cluster ring code (see Fig. \ref{Fig_Codes}), which is capable of correcting an arbitrary error on a single qubit. The resource states for encoding and decoding are obtained by attaching a particle in state $|0_x\rangle$ via $U_{\rm PG}$ operations to all particles of the code state, which is initially prepared in $|0_L\rangle=|C_5\rangle$. This leaves us with $\tfrac{1}{\sqrt{2}}(|0\rangle_A|0_L\rangle_B + |1\rangle_A|1_L\rangle_B)$ \cite{Zw14a}, where $|1_L\rangle = \sigma_z^{\otimes 5} |0_L\rangle$. The encoding and decoding procedure is similar as described for the repetition code, only the required correction operations depending on the outcomes of the Bell measurements differ. In a similar way, one can construct resource states for any CSS code, as codewords are given by stabilizer states, and also encoding, decoding and syndrome-read out can be done by Clifford circuits.

\subsection{Error thresholds}
\label{Sec_ErrorCorrection_threshold}
If one considers noisy resource states, it is straightforward to analyze the performance of the scheme as well as its error thresholds. Consider to this aim the syndrome-readout or error-correction procedure. The noisy resource state is described by Eq. \ref{rho_U}.  If we are interested in protecting quantum information for a longer time, we will apply syndrome-readout and error correction repeatedly. This is done by taking an encoded state, in our example an encoded logical qubit, and a noisy resource state of size $2N$ where Bell measurements are performed on all qubits of the encoded state and the input particles of the resource state. We now consider noise on the input particles and output particles of the resource state separately. The noise on input particle can effectively be moved through the Bell measurements to the encoded state, while noise on output particles acts on the output state and can hence be considered in the subsequent step. This leaves us effectively with a {\em noiseless} resource state that can perform the error correction procedure perfectly. Noise on the input particles, as well as noise on the output particles of the previous step act, via the Bell-measurement, on the encoded input state where the joint effect is described by single-qubit depolarizing noise with parameter $p^2$. We take in addition also some decoherence process on encoded state into account, e.g. because the state is stored for a certain time before error correction is applied, which we again describe by single depolarizing noise with parameter $q$. In total, this leaves us with depolarizing noise with error parameter $p^2q$ that acts on the encoded state before {\em perfect} error correction.

An error correction code can correct a certain number of errors. The threshold for a successful protection is given by the value of the error parameter such that at the logical level, the error is reduced. Since we have shown how to interpret decoherence errors as well as errors from imperfect resource state preparation and imperfect Bell measurements, the threshold for the fault tolerant scheme is simply determined by the error threshold of the underlying error correction code.

For instance, the five-qubit error correction code can correct for one error on one of the five qubits. That is, whenever no or only a single $\sigma_x,\sigma_y$ or $\sigma_z$ error happened, there is no error at the logical level. There might be further error combinations that lead to no error, however we ignore them in order to obtain a simple estimate on the acceptable error rate. We denote by $\tilde p=p^2q$ the total single qubit noise parameter. This corresponds to no error happening with probability $p_{\rm no}=\tfrac{3p+1}{4}$, while with probability $\tfrac{1-p}{4}$ either an $x$,$y$ or $z$ error happens. As all errors are correctible, we can treat them together and just say that an error happens with probability $p_{\rm yes}=\tfrac{3(1-p)}{4}$. At the logical level, we then have no error with probability
\be
p_{\rm no}^{(L)} \geq p_{\rm no}^5 + 5p_{\rm no}^4p_{\rm yes}
\ee
with $p_{\rm no}^{(L)}=\tfrac{3p_L+1}{4}$ and for a successful error correction one needs $p_L \geq p$. From there an error threshold estimate can be determined, and one obtains $\tilde p \geq 0.82517$ \cite{He05}. This translates to an error threshold for the noisy resource state of $p_{\rm crit} = \sqrt[3]{\tilde p} \approx 0.938$ if one assumes $p=q$. Notice that by using a concatenated code, one can achieve $p_L \to 1$ whenever $p> p_{\rm crit}$. An exact treatment of effective noise channels at the logical level, including also concatenated error correction codes, can be found in \cite{Ke13}.
Codes with even higher error thresholds, e.g. Shor type codes, are known \cite{Fe08}. The error threshold for such codes can be as high as $\tilde p= 0.7449$ leading to $p_{\rm crit}=0.9065$. If channel noise is much smaller than noise for resource state preparation and noisy Bell measurements, which can e.g. be assured by using small enough segments in the case of communication, or short time intervals in the case of storage (quantum memory), one can set $q \approx 1$ and obtains an error threshold of $p_{\rm crit} = {\tilde p} \approx 0.8631$, so that up to $13.6 \%$ depolarizing noise per particle can be tolerated.

\section{Encoded quantum communication}
\label{Sec_EncodedQC}

The original scheme for quantum communication based on quantum error correction \cite{KL96} suffered from a very small error threshold (in a gate-based model). Here the measurement-based implementation of error correction offers high error thresholds, which are simply determined by the error threshold of the underlying quantum error correction code as discussed in Sec. \ref{Sec_ErrorCorrection_threshold}. Using large error correction codes, almost 10\% noise per qubit for the resource state and due to the transmission through the channel are acceptable. This is the same order of magnitude as for measurement-based quantum repeaters \cite{Zw12,Zw13}, and thus turns quantum communication based on quantum error correction into a realistic alternative to quantum repeaters. It should be noted that this is also true for a gate-based noise model, which follows from \cite{Kn05}, where error thresholds for fault-tolerant quantum computation using teleportation/measurement-based quantum error correction was used.
Notice that the correction operations can be postponed until the end of the overall channel. Each of the correction operations corresponds to a Pauli operation, and can hence be commuted through the remaining circuit, i.e. does not need to be applied. Only the interpretation of the measurement results and the next required correction operations changes. It is hence sufficient if the measurement outcomes of all stations are sent to the final station, Bob, where the required correction operation is performed.
This offers a significant reduction in the necessary experimental capabilities, as active feed forward is only required at the final station. In addition, one may apply the correction operation on the (unencoded) output state, or not at all if one precesses the state further by means of measurements. This would e.g. be the case if one uses the long-range communication scheme to establish a secret key for cryptography, e.g. using the BB84 \cite{BB84} or the E91 \cite{E91} protocol, where the latter two are based on entangled resource states shared between the communication partners that are subsequently measured. States do not need to be stored in this case, but can be directly measured upon arrival - only the {\em interpretation} of the measurement basis depends on the results of all syndrome measurements during the procedure, and needs to be adjusted appropriately.

Clearly, in a realistic and practical scenario where only small codes are used, the thresholds will be smaller, or the communication distance --in this case specified by the number of error correction steps or equivalently channel segments-- will be limited.

Adjusting the error correction code to the dominant source of channel noise is clearly also beneficial. If only a certain type of error needs to be dealt with, error correction codes with much higher error thresholds are known. For instance, if only dephasing errors (i.e. $Z$ errors) occur, a repetition code is sufficient. In the asymptotic limit of large codewords, such a code offers a protection whenever the probability for no error is larger than $1/2$ \cite{Ke13,Du14,Ar14,Ke14}. This means that for a noise map of the form ${\cal M}(q) = q\rho + (1-q) (\rho + \sigma_z \rho \sigma_z) = p \rho + (1-p)\sigma_z \rho \sigma_z$ for each of the qubits, any $q >0$ suffices and can be corrected by a sufficiently large code. It follows that the threshold value for noisy resource states is also given by $q_{\rm crit} > 0$ or equivalently $p_{\rm crit}>1/2$, which is significantly lower than for general errors. Notice that we have used that noise on each particle is finally described by the map ${\cal M}(q^2\tilde q)$, where $q^2\tilde q >0$ for successful error correction. Here we have moved noise to input state (considering also noise on output states from the previous step), and considered decoherence with error parameter $\tilde q$.

Furthermore, one may also use error avoiding schemes such as encoding into decoherence free subspaces if error are correlated. Another important type of errors are loss errors, in particular when considering quantum communication based on photons through fibers or free space. The performance of measurement-based schemes for direct transmission of encoded information in practical scenarios, i.e. with limited resources or restricted kinds of errors including loss, are currently been investigated and will be reported elsewhere.
A similar approach, following the scheme of \cite{KL96} and using teleportation/measurement-based quantum error correction \cite{Kn05}, was analyzed in \cite{Mu14}, where also loss errors were treated.

To summarize, direct measurement-based quantum communication seems to be a viable alternative to other long-range communication schemes that have been discussed in the literature, including space-based transmission schemes as well as entanglement-based schemes using quantum repeaters. We will discuss measurement-based quantum repeaters, which benefit in a similar way from the measurement-based realization of quantum information processing next.

\begin{figure}[]
\begin{picture}(210,60)
\put(-30,-190){\includegraphics[keepaspectratio, width=12cm]{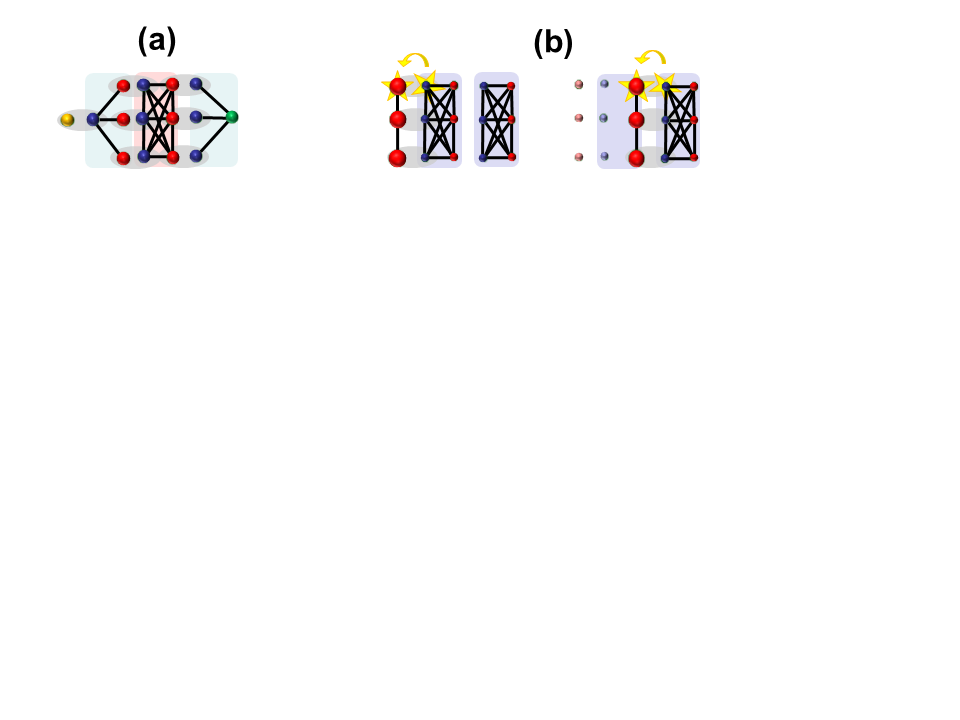}}
\end{picture}
\caption{(a) Resource states for measurement-based communication, including encoding, syndrome-read out (symbolic) and decoding. The input particle is coupled in via a Bell measurement, and encoded information is also processed solely by Bell measurements. (b) Long-distance communication using encoded quantum information and measurement-based error correction. Errors on one side of the Bell measurements can effectively be moved to the other particle.}
\label{Fig_Communication}
\end{figure}

\section{Measurement-based quantum repeaters}
\label{Sec_Repeaters}
\begin{figure}[ht]
\begin{picture}(210,90)
\put(-30,-220){\includegraphics[keepaspectratio, width=14cm]{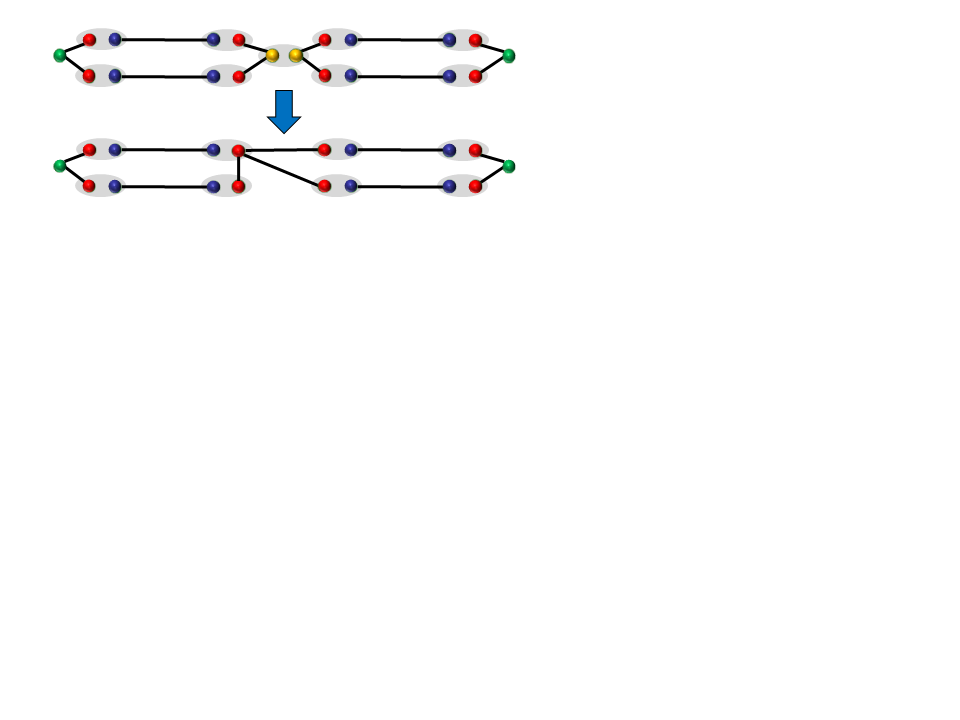}}
\end{picture}
\caption{Illustration of elementary repeater step consisting of entanglement purification of left and right pairs, followed by entanglement swapping. The resource state at the central station can be reduced to contain only input particles.}
\label{Fig_Repeater}
\end{figure}

We now turn to quantum repeaters, where we have described the principal scheme in Sec. \ref{Sec_Repeater}. One of the two building blocks of quantum repeaters, namely entanglement purification, has already been discussed in detail, where the measurement-based implementation offers very high error thresholds of up to $24\%$ depolarizing noise per particle. The second ingredient of a quantum repeater scheme is entanglement swapping, which is performed at intermediate repeater stations. Entanglement swapping corresponds to the teleportation of an input state that is itself entangled, and is realized by performing a Bell measurement. At intermediate repeater stations, elementary pairs are first purified before they are connected by Bell measurements. In principle, one can simply use the resource states for entanglement purification, e.g. the ones for a single or two rounds of the $2 \to 1$ recurrence protocol, specified by resource states of size $2+1$ and $4+1$ respectively (see Fig. \ref{Fig_EPP}). Two such resource states are required for the purification of pairs with the previous and next repeater station, and the output particles are then connected by means of Bell measurements. Again, this can be done beforehand, leaving us with a resource state of reduced size that only contains input particles and no output particles. This resource state performs entanglement purification and connection, and has a size of $N=4$ for a single purification step for each of the pairs, and $N=8$ for two rounds of the recurrence protocol. The corresponding resource states are, up to local unitary operations, graph states, and are depicted in Fig \ref{Fig_Repeater}. The LU operations as well as the interpretation of the Bell measurement outcomes for the success of the purification procedure can be found in \cite{Zw12}.

Notice, however, that the reduced size of the resource state requires that all involved purification steps need to be successful simultaneously. This reduces the achievable rate for the measurement-based repeater. In contrast, a standard implementation might allow for the combination of pairs from previous successful branches, thereby increasing the overall success probability.

This drawback of the measurement-based implementation is compensated by a significantly higher error threshold. Again, the overall repeater scheme offers error thresholds of several percent of acceptable noise per particle, and the more purification steps are realized with a single resource state, the higher is the tolerable noise, similar as in measurement-based entanglement purification \cite{Zw13}. In addition, the usage of $N \to M$ entanglement purification protocols that are not generated from sequential applications of different recurrence round is possible and offers a high yield also in the measurement-based implementation, as illustrated for hashing in Sec. \ref{Sec_Hashing}. In \cite{Zw12}, variants of the repeater scheme that make use of slightly enlarged resource states to offer higher flexibility are discussed.

The ultimate error threshold for a measurement-based quantum repeater can be determined in a similar way as measurement-based entanglement purification. One only needs to consider in addition that at least two pairs need to be connected at the intermediate repeater station. Let us assume that we use an entanglement purification protocol that allows one to distill pairs with unit fidelity in the noiseless case whenever the initial fidelity is larger than $1/2$, which is e.g. the case for multiple rounds of the recurrence protocol. Now it is crucial to note that the resulting particles after the purification of the two channel segments, but prior to connection, are only virtual, i.e. the corresponding resource state that combines entanglement purification and connection does not include them. This implies that {\em no} noise acts on these particles, and one may equivalently describe the situation by resource states for entanglement purification for left and right channel segments, where only input particles are affected by noise but output particles are not. In this case, one can still move noise from noisy resource states to the input pairs from the channel, and the noise of output states in considered afterwards. This leaves us with two perfect pairs that are connected via perfect Bell measurements, i.e. pairs with fidelity $F=1$ on which now the noise on output particles (which we did not consider so far) acts. It follows that the ultimate error threshold for such a measurement-based quantum communication scheme is {\em the same} as for entanglement purification alone. This implies that we have found again a universal and optimal error threshold for quantum communication, where noise up to $24 \%$ per particle can be tolerated. Notice, however, that the transmission rate will be low, as the success probability using such an approach is (exponentially) small. The problem is the non-zero yield for recurrence-type schemes. On the other hand, hashing type entanglement purification protocols offer a non zero yield and hence higher rates, however with slightly smaller error thresholds of about $7\%$ per particle. Notice that there is no issue with memory errors when using hashing protocols, as they only require one-way classical communication, which means that no intermediate processing is required, only at the target station (Bob) the corresponding correction operation is performed.

\section{Hybrid schemes for quantum information processing}
\label{Sec_Hybrid}
We also briefly mention that one can in fact extend the (fault tolerant) measurement-based communication schemes to universal, fault tolerant quantum computation \cite{Zw14a}. We have already seen how to obtain a fault-tolerant quantum memory (or equivalently a communication scheme based on the transmission of encoded states) in a measurement-based way. The corresponding resource states are of minimal size $2N$, where $N$ is the size of the code. Similarly, one can obtain logical Clifford gates with built-in error correction, as outlined in Sec. \ref{Sec_Combination}. The error threshold is the same as for the logical identity operation, i.e. error correction alone. Notice that one can assume $q\approx 1$ in this case, as one can apply error correction repeatedly and hence make external decoherence errors small. In a communication scenario, this would correspond to using very short channels. In this case, the error threshold for resource state is given by the $p_{\rm crit} = \sqrt{\tilde p}$ rather than $p_{\rm crit} = \sqrt[3]{\tilde p}$ when $q=p$ is assumed (see Sec. \ref{Sec_ErrorCorrection_threshold}). The error threshold can be as high as $13.6\%$ depolarizing noise per particle \cite{Zw14a}.
Also logical two-qubit gates, e.g. a CNOT operation, can be constructed in the same way, and with same error threshold.

What is missing for universal quantum computation is the implementation of a certain single-qubit non-Clifford gate, e.g. a $\pi/8$ phase gate. In \cite{Zw14a} we have proposed a measurement-based implementation of such a gate using magic state distillation \cite{Br05} at the logical level. That is, encoded magic states are prepared using fault-tolerantly realized Clifford operations. This allows one to obtain noiseless logical magic states as long as (i) magic state distillation works and (ii) Clifford gates and logical Clifford gates can be implemented fault tolerantly. It turns out that the condition (ii) is more restrictive, i.e. whenever Clifford gates can be implemented fault tolerantly one obtains perfect logical magic states that can be used via gate injection to obtain an encoded $\pi/8$ phase gate. It follows that the same error threshold as for logical Clifford gates or fault-tolerant quantum memories applies, i.e. up to $13.6 \%$ depolarizing noise per particle can be accepted. Even though this number can not be directly compared to gate-based error thresholds, it is nevertheless so high that a measurement-based approach, more precisely a hybrid approach that uses measurement-based elements and combines them with features from the circuit model, is a promising route towards fault tolerance and the practical realization of a universal quantum computer.

Notice that the term hybrid scheme, as used here and in Ref. \cite{Zw14a}, refers to the combination of different models for quantum computation, i.e. elements from the circuit model and from measurement-based quantum computation. The term hybrid quantum information processing or hybrid systems is also often used when dealing with different physical systems, e.g. a combination of discrete and continuous variable systems \cite{VLo14}. This is not what we deal with here.

\section{Experimental realization}
\label{Sec_Experiments}
First experiments that demonstrate important building blocks of measurement-based quantum information processing have already been conducted. Measurement- and teleportation-based single- and two-qubit gates as elements of a one-way quantum computer have been demonstrated with photons \cite{Wa05,Pr07,Ga10} and ions \cite{La13}. Here we discuss recent advances that are more directly related to the measurement-based quantum communication schemes treated here.

\subsection{Experimental realization using photons}
In \cite{Ba14} a proof-of-principle experiment was performed that demonstrates elements of measurement-based quantum error correction. More specifically, an error detection scheme was realized using polarization degrees of freedom of photons. The scheme includes encoding, syndrome-read out and decoding, where a two qubit error detection code that can either detect a phase-flip error on one of the two qubits, or correct for an error if the position of the error is known. To this aim, a four-photon resource state has been prepared. One particle serves as input, one as output, and the intermediate particles hold the encoded quantum information and are subjected to errors. Read-in of information is done by a direct single-qubit measurement on the input particle. This allows one to couple in an (known) quantum state with real coefficients, and prepare an encoded state $\alpha |0_L\rangle + \beta|1_L\rangle$ with $|0_L\rangle =|++\rangle, |1_L\rangle=|--\rangle$. Errors on the intermediate two particles are introduced, and these particles are measured. These measurements perform the decoding to the fourth particle, and also include the error syndrome read-out. The required correction operation can be determined from the measurement outcomes. Notice that also the digitalization of errors was demonstrated. Fidelities of this proof of principle experiment are around $65 \%$.

\subsection{Experimental realization with trapped ions}
A similar experiment to demonstrate all ingredients of measurement-based quantum error correction was performed using trapped ions \cite{La13}. In this case, a full repetition code to correct for phase flip errors was used. In fact, codes of size $3$ and $5$ that are capable of correcting $1$ and $2$ errors respectively were realized. Resource states that allow for encoding, syndrome-read out and decoding that include in addition two particles (one for encoding/read-in, one for decoding/read-out) were deterministically prepared using gate sequences including multi-qubit M{\o}lmer-S{\o}rensen gates. The encoding of quantum information takes place by a measurement of the first (input) particle. Phase errors on the intermediate particles are introduced, and these particles are subsequently measured. These measurements lead to the decoding of quantum information, which is now stored in the last particle. In addition, the outcomes of the measurements reveal the error syndrome and the required correction operation can be determined.

The performance of error correction has been tested in different ways. First, it was demonstrated that one phase-flip error occurring on one of the particles can be fully corrected when using a three-qubit code, and phase-flip errors on two of the particles can be fully corrected when using a 5-qubit code. Second, phase-flip errors occurring on all intermediate particles (i.e. on the encoded state) with a certain error rate were considered, and the fidelity of the decoded particle was determined. This was compared with a direct transmission along a three-qubit chain (one input particle, one intermediate particle where information was not encoded, but which was subjected to phase-flip error with a certain rate, and read-out particle). It was not only demonstrated that error correction is in principle possible for a range of error parameters, but the protection using a three-qubit or five-qubit code does indeed offer an advantage over the direct transmission, as the final fidelity of the output state is larger even when all intermediate particles are subjected to errors. This is the case even though the preparation of larger resource states is clearly more challenging.

A similar experiment could be performed with a different seven-qubit resource state, where a 5-qubit code that is capable of correcting an {\em arbitrary} error on one of the qubits is used. This would allow to demonstrate all elements of full quantum error correction, including non-destructive syndrome read-out.

These experiments show that a measurement-based approach to quantum error correction is not only of theoretical interest, but also offers experimental advantages and has great potential to be an important ingredient in quantum communication and quantum computation schemes.

\section{Other approaches}
\label{Sec_Other}

The first consideration of measurement-based elements in fault-tolerant quantum computation goes back to Knill \cite{Kn05}, where it was shown that with a so-called ancilla-factory approach, one can achieve high error thresholds. The basic idea is to prepare resource states for error correction offline, using entanglement purification as a tool to achieve high fidelities. This results into a probabilistic generation of these states, with a significant overhead. The states are then used to realize error correction in a teleportation-based fashion. Based on numerical analysis, an error threshold for the involved gates of about $1 \%$ per gate was found \cite{Kn05}. Although not explicitly discussed in \cite{Kn05}, this approach can also be applied to quantum communication using direct transmission of encoded quantum information, leading to same error thresholds.

In the context of quantum communication, there are a few related approaches that make use of teleportation/measurement-based quantum information processing. We discuss some of them in the following.

The scheme proposed in \cite{Ji09} uses shared Bell states between repeater station to implement teleportation-based CNOT gates to create encoded Bell states between them, which are finally used for entanglement swapping. A hybrid implementation is discussed in \cite{Ber12}.

The authors of \cite{Mu14} investigate the scheme for quantum communication based on quantum error correction \cite{KL96} combined with teleportation-based quantum error correction \cite{Kn05}. It is thus very similar to \cite{Zw14a}, which was proposed earlier (see also Sec. \ref{Sec_EncodedQC}). An optical implementation is considered in \cite{Ew15}.

An interesting approach based on time reversal of standard quantum repeaters \cite{Br98} was put forward in \cite{Az15}. Here the entangling operation for entanglement swapping at each repeater station is performed before the creation of entanglement between these stations. In order to make this robust against loss and and other imperfections one uses specially tailored photonic graph states which are processed in a measurement-based way. This scheme avoids the need for quantum memories and requires fast feed-forward only within each repeater station.

We remark that a large variety of different proposals for long-range quantum communication schemes have been put forward in recent years by many groups, and also a number of experiments have been performed that demonstrate important building blocks of long-range quantum communication schemes. Our aim is not to provide an overview on all these works. In this paper we have rather concentrated on schemes where measurement-based elements play a crucial role.

\section{Summary and outlook}
\label{Sec_Summary}
In this review article, we have given an overview over some recent approaches to use measurement-based elements in quantum communication. We have considered direct communication of encoded quantum information, where encoding, decoding as well as error correction are performed in a measurement-based way. Input states are coupled via Bell-measurements to resource states, and apart from channel noise the only source of imperfections are noisy resource states and noisy Bell measurements. Error thresholds for a fault-tolerant implementation are solely determined by error thresholds for the underlying error correction codes, and can be as high as $13\%$ per qubit.

Similarly, measurement-based elements can be used in an entanglement-based communication scheme using quantum repeaters. There, entanglement swapping and entanglement purification are the central building blocks, which can be realized and combined in a measurement-based way. The corresponding resource states are of minimal size, containing only input and output particles. In this case, even larger error thresholds can be found, which are in addition optimal and universal. For entanglement purification as well as for entanglement-based quantum communication, up to $24 \%$ of noise per particle can be accepted.

It should be emphasized that these error threshold correspond to an error model that is reasonable, conservative, and consistent with the primitives of measurement-based quantum information processing. These numbers can, however, not be directly compared to thresholds obtained for gate based error models. Working out such a connection is an interesting open problem. In addition, an extension of the analysis to correlated errors would be interesting, where the localization of errors and hence a justification of a local error model using multipartite entanglement purification for resource states is a promising approach. Finally, we mention that the measurement-based realization of quantum communication also allows for a simple security proof, by showing that resulting entangled pairs and hence the resulting secret key is in fact private. This will be reported elsewhere \cite{Zw15}.

Long-distance quantum communication remains an important and challenging goal. The use of measurement-based elements may turn out to be a key ingredient towards a practical realization, as experimental and practical requirements can be significantly relaxed.

\textit{Acknowledgements.---}
This work was supported by the Austrian Science Fund (FWF): P24273-N16, P28000-N27, SFB F40-FoQus F4012-N16.


\begin{thebibliography}{99}

\bibitem{Sca09}
V. Scarani, H. Bechmann-Pasquinucci, N. J. Cerf, M. Du\u{s}ek, N. L\"utkenhaus, and M. Peev, Rev. Mod. Phys. {\bf81}, 1301 (2009).

\bibitem{Bro09}
A. Broadbent, J. Fitzsimons, E. Kashefi, in {\em Proceedings of the 50th Annual Symposium on Foundations of Computer
Science} (IEEE Computer Society, Los Alamitos, CA, 2009), pp. 517.

\bibitem{Bar12}
Stefanie Barz et al., Science 335, 303 (2012).

\bibitem{Kimble08}
H. J. Kimble, Nature {\bf453}, 1023 (2008).

\bibitem{KL96}
E. Knill and R. Laflamme, arXiv:quanth-ph/9608012 (1996).

\bibitem{Br98}
H. J. Briegel, W. D\"{u}r, J. I. Cirac, and P. Zoller, Phys. Rev. Lett. {\bf81}, 5932 (1998).

\bibitem{Be93}
C. H. Bennett {\em et al.}, Phys. Rev. Lett. {\bf70}, 1895 (1993).

\bibitem{Gi11}
N. Sangouard, C. Simon, H. de Riedmatten, and N. Gisin, Rev. Mod. Phys. {\bf83}, 33 (2011).

\bibitem{Je13}
T. Jennewein and B. Higgins, Quantum space race, Physics World 26(3), 5256, (March 2013).

\bibitem{Sho95}
P. Shor, Phys. Rev. A {\bf52}, 2493 (1995).

\bibitem{Ste96}
A. M. Steane, Phys. Rev. Lett. {\bf77}, 793 (1996).

\bibitem{Go97}
D. Gottesman, {\emph{Stabilizer codes and quantum error correction}}, PhD thesis, Caltech (1997). arXiv:quant-ph/9705052



\bibitem{Zu93}
M. Zukowski, A. Zeilinger, M. A. Horne, and A. K. Ekert, Phys. Rev. Lett. {\bf71}, 4287 (1993).


\bibitem{Be96}
C. H. Bennett, G. Brassard, S. Popescu, B. Schumacher, J. A. Smolin, and W. K. Wootters, Phys. Rev. Lett. {\bf76}, 722 (1996).

\bibitem{De96}
D. Deutsch, A. Ekert, R. Jozsa, C. Macchiavello, S. Popescu, and A. Sanpera, Phys. Rev. Lett. {\bf77}, 2818 (1996).

\bibitem{Du07}
W. D\"{u}r and H. J. Briegel, Rep. Prog. Phys. {\bf70}, 1381 (2007).

\bibitem{Ra01}
R. Raussendorf and H. J. Briegel, Phys. Rev. Lett. {\bf 86}, 5188 (2001).

\bibitem{Ra01b}
H. J. Briegel and R. Raussendorf, Phys. Rev. Lett. {\bf 86}, 910 (2001).

\bibitem{Zw12}
M. Zwerger, W. D\"ur, and H. J. Briegel, Phys. Rev. A {\bf85}, 062326 (2012).

\bibitem{Zw13}
M. Zwerger, H. J. Briegel, and W. D\"ur, Phys. Rev. Lett. {\bf110}, 260503 (2013).

\bibitem{Zw14a}
M. Zwerger, H. J. Briegel, and W. D\"ur, Sci. Rep. {\bf4}, 5364 (2014).

\bibitem{La13}
B. P. Lanyon, P. Jurcevic, M. Zwerger, C. Hempel, E. A. Martinez, W. D\"ur, H. J. Briegel, R. Blatt, and C. F. Roos, Phys. Rev. Lett. {\bf111}, 210501 (2013).

\bibitem{Ba14}
S. Barz, R. Vasconcelos, C. Greganti, M. Zwerger, W. D\"ur, H. J. Briegel, and P. Walther, Phys. Rev. A {\bf90}, 042302 (2014).

\bibitem{Zw14b}
M. Zwerger, H. J. Briegel, and W. D\"ur, Phys. Rev. A {\bf90}, 012314 (2014).

\bibitem{Ja72}
A. Jamiolkowski, Rep. Math. Phys. {\bf 3}, 275-278 (1972).

\bibitem{Ch75}
M. D. Choi, Linear Alg. and Its Appl. {\bf10}, 285-290, (1975).

\bibitem{NiCh98}
Michael A. Nielsen und Isaac L. Chuang, {\em Quantum Computation and Quantum Information}, Cambridge University Press  (2000).

\bibitem{Kn05}
E. Knill, Nature {\bf434}, 39 (2005).

\bibitem{Rau07}
R. Raussendorf and J. Harrington, Phys. Rev. Lett. 98, 190504 (2007).


\bibitem{He06}
M. Hein {\em et al.}, In \emph{Proceedings of the International School of
Physics ``Enrico Fermi'' on ``Quantum Computers, Algorithms and
Chaos''} (2005); arXiv:quant-ph/0602096.


\bibitem{Go09}
D. Gottesman, arXiv:0904.2557, (2009).

\bibitem{Ra12}
R. Raussendorf, Phil. Trans. R. Soc. A {\bf370}, 4541 (2012).

\bibitem{Br09}
H. J. Briegel, D. E. Browne, W. D\"ur, R. Raussendorf and M. Van den Nest, Nature Physics {\bf5}, 19 (2009).

\bibitem{Be96a}
C. H. Bennett, D. P. DiVincenzo, J. A. Smolin, and W. K. Wootters, Phys. Rev. A 5{\bf4}, 3824 (1996).

\bibitem{As04}
H. Aschauer, \emph{Quantum communication in noisy environments}, Dissertation, LMU M\"unchen (2004).

\bibitem{Ha07}
L. Hartmann, B. Kraus, H. J. Briegel, and W. D\"ur, Phys. Rev. A {\bf 75}, 032310 (2007).

\bibitem{Nie04}
M. A. Nielsen, Phys. Rev. Lett. 93, 040503 (2004).

\bibitem{ProbGates}
S. D. Barrett and P. Kok,
Physical Review A \textbf{71}, 060310 (2005);
L. M. Duan and R. Raussendorf,
Phys. Rev. Lett. \textbf{95}, 080503 (2005);
D. Gross, K. Kieling and J. Eisert,
Physical Review A \textbf{74}, 042343 (2006);
K. Kieling, T. Rudolph and J. Eisert,
Phys. Rev. Lett. {\textbf 99}, 130501 (2007);
D.E. Browne and T. Rudolph,
Phys. Rev. Lett. \textbf{95}, 010501 (2005).




\bibitem{Ra03}
R. Raussendorf, D. E. Browne and H. J. Briegel, Phys. Rev. A {\bf 68}, 022312 (2003).

\bibitem{He04}
M. Hein, J. Eisert, and H. J. Briegel, Phys. Rev. A {\bf69}, 062311 (2004).

\bibitem{MS99}
A. S{\o}rensen and K. M{\o}lmer, Phys. Rev. Lett. {82}, 1971 (1999).

\bibitem{Di08}
S. Diehl, A. Micheli, A. Kantian, B. Kraus, H. Büchler, P. Zoller, 
Nature Physics 4, 878 (2008).

\bibitem{Ve09}
F. Verstraete, M. M. Wolf and J. I. Cirac, 
Nature Physics 5, 633 (2009).

\bibitem{Du03}
W. D\"ur, H. Aschauer and H. J. Briegel, Phys. Rev. Lett. {\bf91}, 107903 (2003).


\bibitem{Gl06}
S. Glancy, E. Knill, and H. M. Vasconcelos, Phys. Rev. A {\bf74}, 032319 (2006).

\bibitem{Kr06}
C. Kruszynska, A. Miyake, H. J. Briegel, W. D\"ur, Phys. Rev. A {\bf74}, 052316 (2006).

\bibitem{Wa15}
J. Waln\"ofer and W. D\"ur, in preparation.



\bibitem{He05}
M. Hein, W. D\"ur, and H. J. Briegel, Phys. Rev. A {\bf71}, 032350 (2005).

\bibitem{Ke13}
F. Kesting, F. Fr\"owis, and W. D\"ur, Phys. Rev. A {\bf88}, 042305 (2013).

\bibitem{Fe08}
J. Fern and K. B. Whaley, Phys. Rev. A {78}, 062335 (2008).


\bibitem{BB84}
C. H. Bennett and G. Brassard, Proceedings of IEEE International Conference on Computers, Systems and Signal Processing, Volume 175, page 8 (1984).

\bibitem{E91}
A. K. Ekert, Phys. Rev. Lett. {\bf67}, 661 (1991).


\bibitem{Du14}
W. D\"ur, M. Skotiniotis, F. Fr\"owis, and B. Kraus, Phys. Rev. Lett. {\bf112}, 080801 (2014).


\bibitem{Ar14}
G. Arrad, Y. Vinkler, D. Aharonov, and A. Retzker, Phys. Rev. Lett. {\bf112}, 150801 (2014).

\bibitem{Ke14}
E. M. Kessler, I. Lovchinsky, A. O. Sushkov, and M. D. Lukin, Phys. Rev. Lett. {\bf112}, 150802 (2014).

\bibitem{Mu14}
S. Muralidharan, J. Kim, N. L\"utkenhaus, M. D. Lukin, and L. Jiang, Phys. Rev. Lett. {\bf112}, 250501 (2014).

\bibitem{Br05}
S. Bravyi and A. Kitaev, Phys. Rev. A {\bf71}, 022316 (2005).

\bibitem{Wa05}
P. Walther, K. J. Resch, T. Rudolph, E. Schenck, H. Weinfurter, V. Vedral, M. Aspelmeyer, and A. Zeilinger, Nature {\bf434}, 169 (2005).

\bibitem{Pr07}
R. Prevedel, P. Walther, F. Tiefenbacher, P. B\"ohi, R. Kaltenbaek, T. Jennewein, and A. Zeilinger, Nature {\bf445}, 65 (2007).

\bibitem{Ga10}
W.-B. Gao, A. M. Goebel, C.-Y. Lu, H.-N. Dai, C. Wagenknecht, Q. Zhang, B. Zhao, C.-Z. Peng, Z.-B. Chen, Y.-A. Chen, J.-W. Pan, PNAS 2010 {\bf107} (49) 20869-20874 (2010).


\bibitem{Ji09}
L. Jiang, J. M. Taylor, K. Nemoto, W. J. Munro, R. Van Meter, M. D. Lukin, Phys. Rev. A {\bf79}, 032325 (2009).

\bibitem{Ber12}
N. K. Bernardes, and P. van Loock, Phys. Rev. A {\bf86}, 052301 (2012).


\bibitem{Ew15}
F. Ewert, M. Bergmann, and P. van Loock, arXiv:1503.06777 (2015).

\bibitem{Az15}
K. Azuma, K. Tamaki, and H.-K. Lo, Nature Communications {\bf6}, 6787 (2015).

\bibitem{VLo14}
U. L. Andersen, J. S. Neergaard-Nielsen, P. van Loock, A. Furusawa, E-print arXiv:1409.3719 .

\bibitem{Zw15}
M. Zwerger et al., in preparation


%
%
%
%
%
%
%
%
%
%
%
%
%
%
%
%
%
%
%
%
%





\end{thebibliography}
\end{document}